# High-performance super-capacity using activated glassy carbon foam

Xiaoqian Wang, Mingming Ma

University of Science and Technology of China, Hefei, China, 230026

Abstract: An urgent demand for new sustainable and efficient energy conversion and storage devices required the development of novel electrode materials with increased specific capacitance. However, low mass loading, poor scalability, and low working voltage always limited further practical application of most reported high-performance supercapacitors electrode materials. In this work, we demonstrated the preparation of activated glassy carbon foam as high-performance electrode via oxidizing commercial glassy carbon foam in HNO3 solution. Because of its internal self-supporting porous structure with high conductivity and excellent chemical stability, the single electrode delivers both ultrahigh specific capacitance of 678.6 F·g-1 and absolute capacitance of 11.3 F at a wide potential window of 0 – 1.6V in 1M H2SO4 solution with no need for extra current collector and substrate. Furthermore, we assemble a symmetric supercapacitor device with activated glassy carbon foam, which could endure a working voltage of 1.6 V in 3M H2SO4 and reached a high energy density of 54.3 Wh·kg-1 at a power density of 38.5 KW·kg-1, while possessing an impressive absolute power of 0.65 W. Significantly, there's no any decay of capacitive performance after 10000 cycles. Therefore, the present work could open up a new way to future exploration and large-scale production of high-performance electrode materials for supercapacitors devices.

Introduction

Climate change and depletion of fossil fuels have been greatly affecting the global environment and economy. Meanwhile, portable electronic devices and the hybrid electric vehicles continue to break new market around the world. As a result, there has been an urgent and ever-increasing demand for sustainable and environmentally friendly power sources as well as clean and efficient devices for energy conversion and storage. Among various choices, electrochemical capacitors (ECs), which also known as supercapacitors (SCs), are in the spotlight for their excellent performance[1]. Compared with batteries, supercapacitors could provide great higher power density and exhibit a much more excellent long-term stability, making them the hopeful energy-storage devices on numerous occasions. However, attributed to their relatively low energy density, the practical applications of supercapacitors have always been limited. To meet the urgent requirements of society, appreciable efforts have been devoted towards improving the value of specific capacitance of SCs through various ways during the past few decades. Unfortunately, most of those materials reported owning high specific capacitance are far away from the commercial applications owing to their poor scalability and rather low load of activated materials. Because the interface area for the surface reactions, which are the base of charge storage in the SCs, doesn't increase proportionally when the mass of electrodes rises, those devices always exhibit poor specific capacitance at relatively large scale.

During the exploration of novel electrode materials with high specific capacitance and electrolytes with wider potential windows, carbon-based materials have attracted tremendous attention, such as activated carbon[2-3], porous carbon[4], carbon aerogels[5], carbon cloths[6], graphene[7] and carbon nanotubes (CNTs)[8-9]. Unlike other super stars among carbon materials like graphene and CNTs, glassy carbon foam (GCF) has kept a low profile for a long time. However, as the competing for alternative materials of high-performance SCs continues to heat up, GCF shows more and more superiority in this competition, making it move into the spotlight. First of all, the self-supporting sponge structure with high conductivity of GCF gives it the ability to serve as both activated and conducting material. In other words, there's no need to introduce extra current collectors to the SCs devices, which will significantly boost the properties of the whole device. Secondly, GCF exhibits a rather stable chemical property so that the devices could endure acid electrolyte of higher concentration and a wider potential window in aqueous systems. On one hand, $H^+$ holds the highest charge density, so the electrical double-layer capacitance could reach a higher value in acid solutions than in ionic liquids or organic electrolytes, and on the other hand, the potential window could improve the energy density quadratically. Last but not least, the commercial GCF could reach relatively large scale, making it possible to get supercapacitor devices at more practical scale. Compared with activated carbon materials used in commercial SCs, the activated glassy carbon foam (AGcf) we got exhibits much higher specific capacitance. While being compared with the high-performance electrode materials like graphene-based and CNTs-based materials, the AGCF supercapacitor devices could reach much higher absolute capacitance. All these superiorities mentioned above optimize the overall performance of supercapacitors devices, which paves the way towards future application.

Our strategy to improve the electrochemical performance of commercial GCF is to increase its surface area for boosting the electrical double-layer capacitance while introducing oxygen-containing and nitrogen-containing functional groups onto its surface for producing pseudocapacitance simultaneously by oxidizing commercial glassy carbon foam in $HNO_3$ solution at 90 ℃ for 8 h (See the supporting information). Figure 1a&b shows that the untreated Gcf has ultra-smooth surface. After oxidative activation, the surface of the samples became relatively rough (Figure 1c&d), indicating the successful modification of the glassy carbon foam surface by oxidation in $HNO_3$.

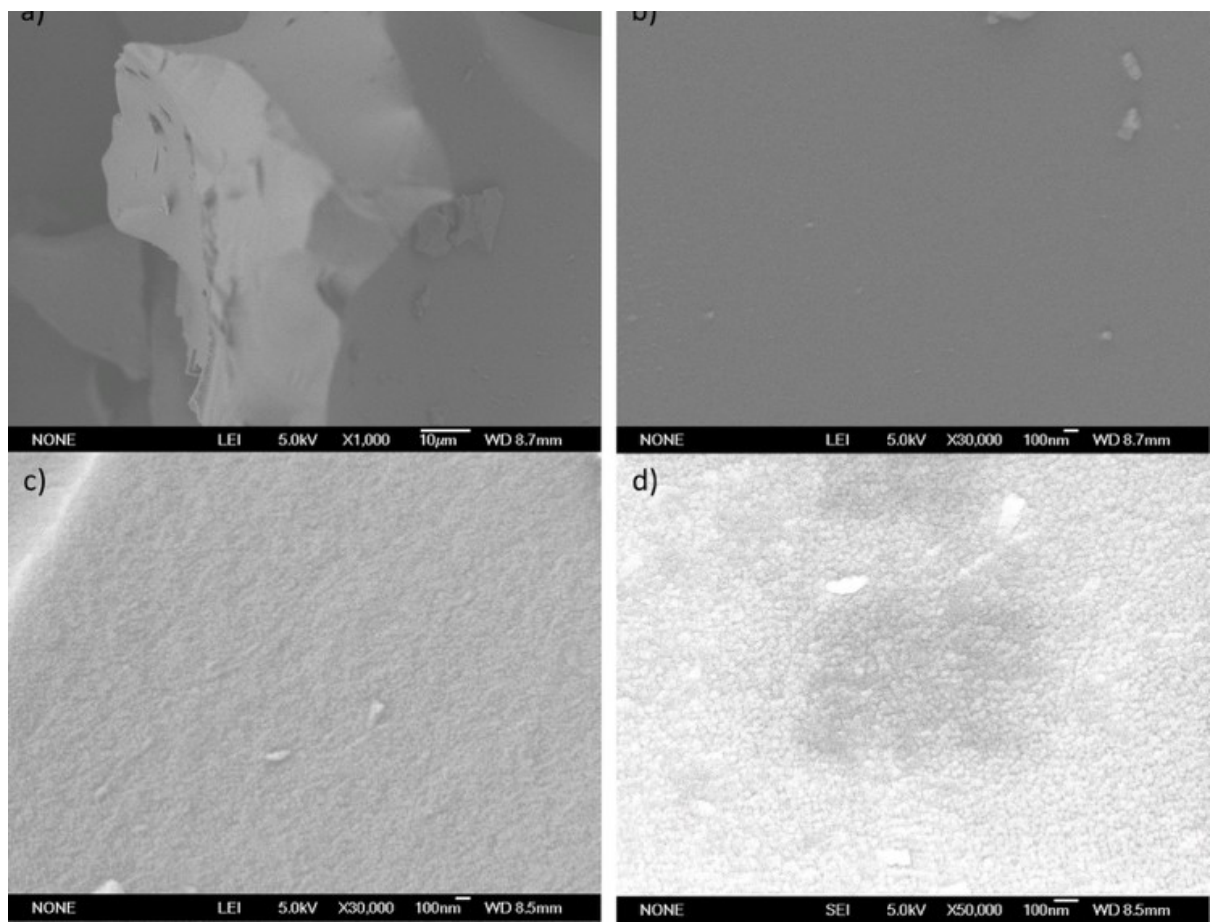

Fig 1. a) and b) SEM images of untreated glassy carbon foam. c) and d) SEM images of AGcf08.

To further investigate the influence on electrochemical performance of erosion by $HNO_3$, $N_2$ adsorption–desorption isotherms analysis was carried out (see the supporting information). The results reveal that the Brunauer–Emmett–Teller (BET) surface area of glassy carbon foam significantly increased from 0.5925 $m^2/g$ to 3.5389 $m^2/g$ upon oxidative activation in $HNO_3$ solution. Other than specific surface area, the pore size distribution also plays an important role in the electrochemical performance. According to the previous reported work[10], the micropores, especially those smaller than 1 nm, will lead to surprisingly high capacitance, the mesopores with a range of 2 – 50 nm provide diffusion channels for ions (important for high power output), and the uniform mesopores of carbon materials could result in high rate capability[11]. It is clearly shown that in Figure1 supporting information, compared with the pore size distribution of untreated Gcf, AGcf-8 shows not only a wider range of pore size distribution but also a significantly increase in the amount of micropores and mesopores. As a result of dramatically increased specific surface area and more excellent pore size distribution, AGcf-8 electrode could exhibit an ultra-high electric double-layer capacitance while maintaining a high power density as well as outstanding rate capability.

As is known to all, the pseudo-capacitance mainly depends on the surface faradaic reactions of electrode materials. Therefore, the structure and valence state of surface atoms were also studied. The overall XPS spectrum of untreated glassy carbon foam and AGcf-8 were compared in Figure 2a. It can be seen that C 1s peak in AGcf-8 decreased while N 1s and O 1s peaks were enhanced, indicating that the increase in nitrogen and oxygen content. In comparison to the untreated Gcf, the AGcf-8 sample exhibited a broader shoulder peak in C 1s spectra, which reflects the presence of four types of carbon bonds (Fig 2b): C–C/C=C (284.4 eV), C-O (286.2 eV), C=O (287.3 eV), and O-C=O (288.9 eV), indicating the significantly increased functional groups on the glassy carbon foam surface after oxidative activation in $HNO_3$. The broad peak in O 1s core-level spectrum of the AGcf-8 was also be deconvoluted into peaks corresponding to >C=O, >COH, and –COOH groups (Figure 2c), which is consistent with the C 1s result. The oxygen ratio (O/(O + C), representing atoms number) calculated from XPS spectrum increased from 2.1% (untreated Gcf) to 15.5% (AGcf-8), further confirming the successful activation. In addition, Raman spectra presented the D:G peak intensity ratio of AGCF-8 (0.94) sample was closed to that of untreated GCF (0.99) (Figure 2C), indicating the carbon atoms in the glassy carbon foam hold an ultra-stable structure. It is well known that doping with some heteroatoms like N and O, could not only boost the bulk electric but also enhance the wettability of carbon frameworks, resulting better utilization of the large specific surface area of the carbon materials

Cyclic voltammetry (CV) and galvanostatic charge-discharge (GCD) were used to investigate the performance of the AGcf-8 as an electrode material for SCs in a three-electrode configuration with 1.0 M H2SO4 as electrolyte. CV and GCD curves of the untreated Gcf and AGcf-8 electrodes are collected in Figure 3a,b. Remarkably, the CV curve of the AGcf-8 electrode exhibits a box-like shape with good symmetry at a sufficiently wide voltage window (0 – 1.6 V), suggesting the AGcf-8 electrode can be able to operate at 0 to 1.6 V. Such the large overpotential of HER and OER should be attributed to the excellent stability of the glassy carbon foam. Notably, the areal capacitance of AGcf-8 electrode calculated from the discharge at 10 mA·cm-2 achieved 22621.4 mF·cm-2 (678.6 F·g-1), which makes an 75404-fold enhancement compared to the untreated Gcf electrode (0.3 mF·cm-2) and is much higher than the values of previously reported carbon-based materials and even some metal oxides electrodes, such as EACC-10[6], $WO_{3-x}/MoO_{3-x}$[12], $Ni/Co_3O$[13], and chemically activated carbon cloth[3]. Electrochemical impedance spectroscopy (EIS) was also conducted to characterize the electrochemical properties of the AGcf-8 electrode, as shown in Figure 3c. The plot displays a semicircle in the high frequency region representing the contact resistance of as small as 2.03Ω and charge transfer resistance of 5.34Ω, while demonstrating a nearly ideal capacitive behaviour with a vertical slope at the low frequency region. Additionally, Figure 3d shows that the AGcf-8 electrode possessed a prominent cycling stability with no any decay of capacitance after 10000 cycles.

Moreover, the AGcf electrodes exhibited an outstanding rate capability with increased current density, which retain more than 87% of the areal capacitance as the current density increased from 5 to 20 mA·cm-2, suggesting its great potential in practical application.

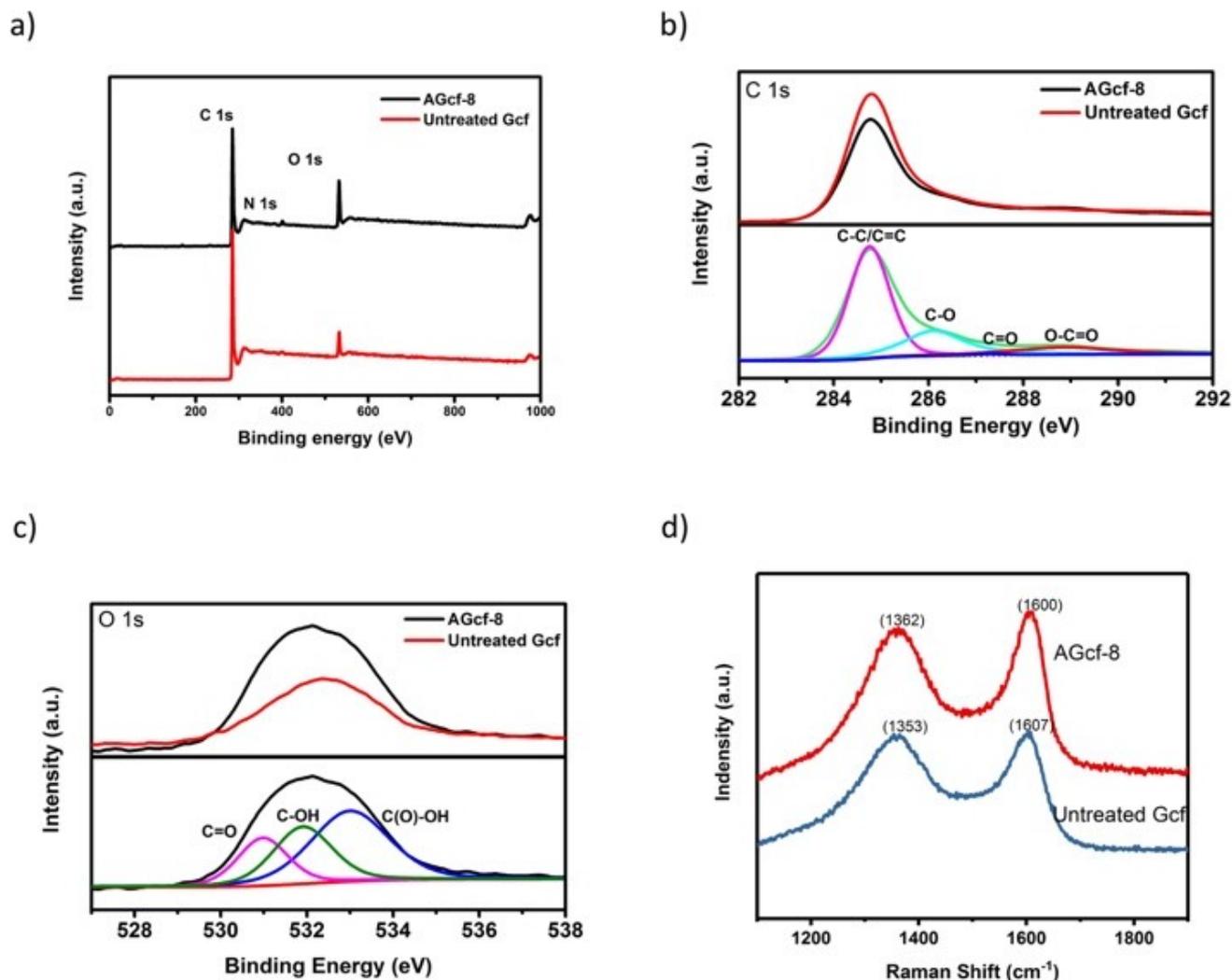

Fig 2. a), b) and c) Overall, core C 1s and core O 1s XPS spectra of untreated Gcf and AGcf-8. d) Raman spectra of the samples.

Figure 3e compares the areal capacitance and rate performance of AGcf electrodes oxidated in different time with other reported electrode materials, indicating 8h is an appropriate oxidating time. Significantly, the absolute capacitance of a single AGcf-8 electrode achieved as high as 11.3 F while possessing a high specific capacitance as 678.6 F·g-1. Such a tremendous value is unparalleled among nearly all the high-performance electrode materials reported. We compared the absolute capacitance and specific capacitance of the AGcf-8 electrode with other electrode materials fabricated in lab and commercial supercapacitors, such as core–shell ultramicroporous carbon nanospheres[14], porous carbon with small mesopores[15], 3D hierarchical porous carbon[16], F, N co-doped carbon microspheres[17], and N-doped mesoporous carbon[18], as shown in Figure 3f.

Two-electrode symmetric supercapacitors have been also assembled with 3M $H_2SO_4$ as electrolyte and PVA/$H_3PO_4$ membrane[19] as separator, to test the feasibility of the AGcf-8 as high performance electrodes for symmetric SCs. Low mass loading is always Achilles' heel for most reported electrode materials, however, thanks to the initial high conductivity and self-supporting sponge structure of AGcf-8, we can get relatively large-scale devices with no need for current collector and substrate. So the electrochemical performance of the whole SCs devices makes more sense than others.

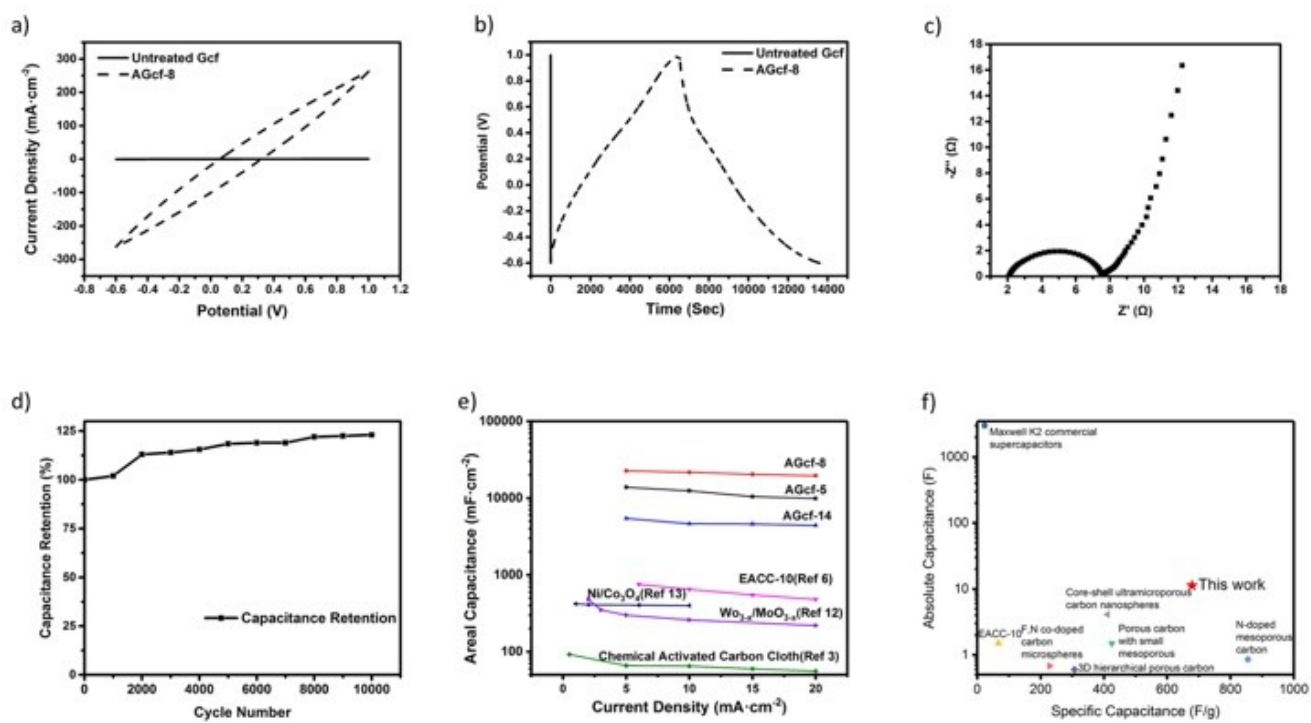

Fig 3. a) and b) Comparison of the CV and GCD curves of untreated Gcf and AGcf-8 electrodes. c) Nyquist plot of AGcf-8 electrode. d) Capacitance retention of AGcf-8 electrode at 10 mA·cm-2. e) Areal capacitance as a function of current density of several different materials for SCs electrodes. f) Excellent performance of AGcf-8 in the competition of both specific and absolute capacitance.

Figure 4a presents the CV curves of the symmetric SC device for the voltage of from 0 to 1.6 V with scan rate varying from 1 to 40 mV·s-1. GCD curves at different current density from 5 to 20 mA·cm-2 are collected in Figure 4b. The specific capacitances calculated from the discharge curves reached 152.7, 145.1, 136.2 F·g-1 at different current densities of 5, 10, 20 mA·cm-2, suggesting an outstanding rate capability of the device. EIS of the frequency range from 0.01 Hz to 100 KHz yields the Nyquist plot shown in Fig. 4C. It is clearly shown that there're two semicircles in the plot, which represent the electrical double layer capacitance and pseudo-capacitance respectively.

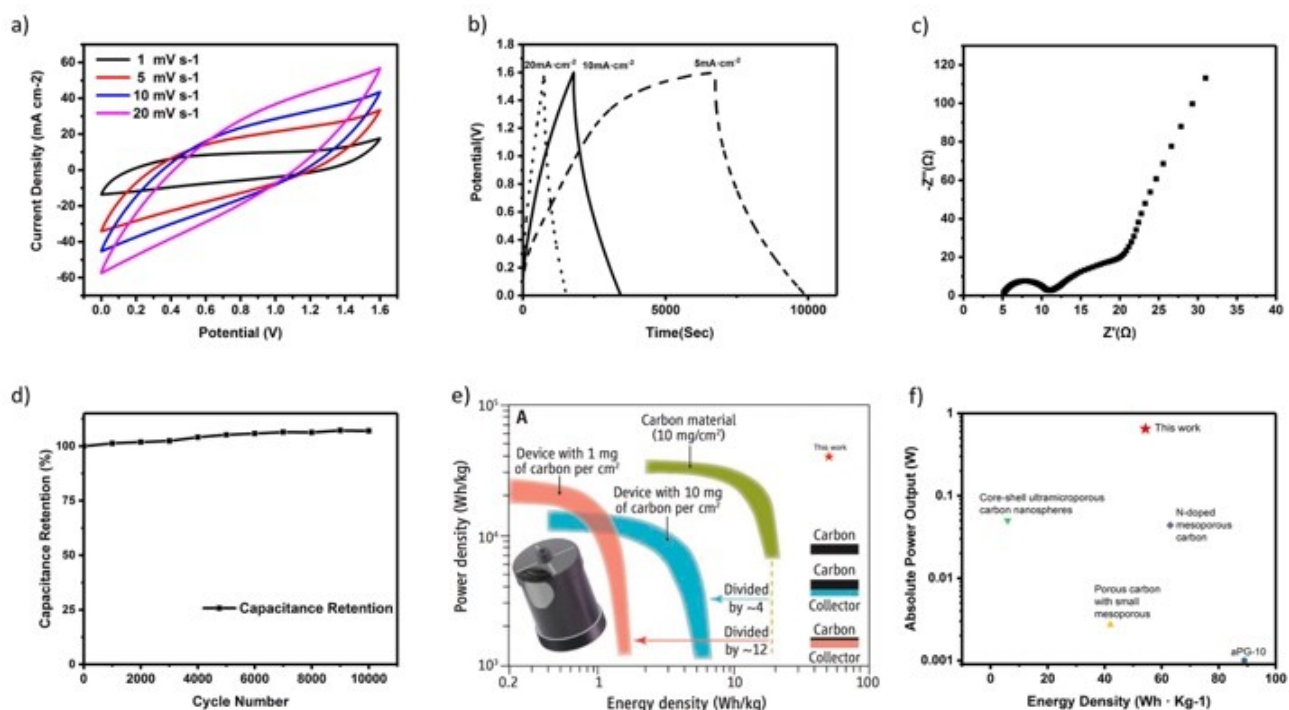

Fig 4. a) and b) CV and GCD curves of AGcf-8 symmetric SC device at different scan rate and different current density. c) EIS plot of the symmetric SC device. d) Long-term cycling stability of the symmetric SC device at 10 mA·cm-2. e) Ragone plot of carbon-based SCs devices, from Ref 20. f) Unparalleled absolute power output of AGcf-8 symmetric SC device compared with others reported.

To further evaluate the its suitability for practical applications, the long-term cycling stability and reversibility of the AGcf-8 symmetric SC device have been also studied. The device exhibited excellent cycling stability with no any capacitance decrease after 10000 cycles, as shown in Figure 4d, which is rather important for the practical utilizations. The energy storage performance of the AGcf-8 symmetric SC device is shown in Figure 4f[20], by comparing with the values reported for other SC devices based on carbon materials. Attributing to the broad operating voltage (up to 1.6 V) and excellent specific capacitance (152.7 F·g-1) at aqueous system, the AGcf-8 symmetric cell reached a high energy density of 54.3 Wh·kg-1 at a power density of 38.5 KW·kg-1. Moreover, the absolute power output of the AGcf-8 symmetric SC device reached an amazing value of 0.65W, which was far ahead of other high energy density SCs devices reported, such as aPG -10[21], core–shell ultramicroporous carbon nanospheres[14], porous carbon with small mesopores[15], and N-doped mesoporous carbon[18], as shown in Fig 4e.

In summary, we have prepared excellent supercapacitors electrodes materials from commercial glassy carbon foam through an efficient an effortless HNO3 oxidation method. During the oxidation, the smooth surface of GCF became rough and abundant functional groups present onto the surface of the sample. On the one hand, the increased surface area and improved wettability enhanced the electrical double-layer capacitance dramatically. On the other hand, the presence of oxygen-containing and nitrogen-containing functional groups boosted the pseudo-capacitance significantly. The AGCF-8 electrode reached a high areal capacitance of 22621.4 mF·cm-2 at a high current density of 5 mA·cm-2. Significantly, the absolute capacitance of the electrode reached an impressive value of 11.3 F. Moreover, a symmetric SC device with an extraordinary energy density of 54.3 Wh·kg-1 and remarkable working voltage of 1.6 V in aqueous systems was achieved by using the AGCF-8 as electrodes. The device held a significant absolute power at 0.65 W and exhibited excellent long-term durability with no any decay of capacitive performance after 10000 cycles. The methodology developed in this study provides us a brand new way to high-performance, stable-property, and large-scale production of supercapacitors electrode materials with wider working voltage in aqueous systems, which holds a great promise for future application of practical high-performance electrochemical supercapacitors.

# Experiment Section

## Chemicals and Materials

Polyvinyl alcohol (PVA, 99% hydrolyzed, degree of polymerization 1750), nitric acid ($HNO_3$) and sulfuric acid ($H_2SO_4$), phosphoric acid($H_3PO_4$) were purchased from Shanghai Chemical Reagent Co. Ltd. All chemicals were of analytical grade and were directly used without any purification. Glassy carbon foam was purchased from Alfa Aesar Co. Ltd.

## Preparation of Activation of Glassy carbon foam

Typically, 10 mL 65% $HNO_3$ were put into a glass vial (20 mL) and then a bar of glassy carbon foam was immersed into it. The vial was maintained at 90℃ for different time (4, 6, 8h). The activated glassy carbon foam was taken out and washed with deionized water before dried at 60℃.

## Characterization of Activated Glassy carbon foam (AGcf)

xxx

## Electrochemical characterization of AGcf

The electrochemical tests of single AGcf electrodes were performed in a typical three-electrode system set up with an electrolyte solution of 1M $H_2SO_4$, using AGcf as the working electrode, a stainless steel plate as the counter electrode and Ag/AgCl as the reference electrode. The electrochemical workstation is CHI 660E from Chenhua Shanghai. Cyclic voltammogram (CV) tests were performed in the potential range of 0 - 0.7 *V* vs. Ag/AgCl under a scan rate of 10 - 100 $mV·s^{-1}$. Galvanostatic charge–discharge (GCD) tests were performed by scanning from 0 to 0.7 *V* at current densities from 0.1 – 20 $mA·cm^{-2}$. Electrochemical impedance spectroscopy (EIS) tests were done with a frequency range of 100 *kHz* - 0.01 *Hz* at a 5 *mV* amplitude referring to open circuit potential.

## Preparation of H$_3$PO$_4$/PVA/H$_2$O

Mixing polyvinyl alcohol (PVA) (99% hydrolyzed, degree of polymerization 1750) powder with water (1 g PVA/10 mL H$_2$O) and subsequent heating under stirring to ~90°C until the solution becomes clear. After cooling down, concentrated phosphoric acid(0.8 g H$_3$PO$_4$) was added and the viscous solution was stirred thoroughly. Finally, the clear solution was cast into a Petri dish where it was left to let excess water evaporate. Once the polymer electrolyte (H$_3$PO$_4$/PVA) became hard, it was cut into pieces matching the size of the electrodes.

## Fabrication of Aqueous symmetric supercapacitor (H$_2$SO$_4$ as electrolyte)

The as-prepared AGcf electrodes were immersed in the 1M H$_2$SO$_4$ solution for 10 h to ensure them to be completely wetted. A piece of polycarbonate membrane (serving as the separator) was saturated with 1M H$_2$SO$_4$ electrolyte. Two AGcf electrodes (X cm×X cm) and the polycarbonate membrane were sandwiched together to get the symmetric supercapacitor.

## Fabrication of Aqueous symmetric supercapacitor (BMIM-NTF$_2$ as electrolyte)

The as-prepared AGcf electrodes were immersed in the BMIM-NTF$_2$ solution for 10 h to ensure them to be completely wetted. A piece of PVA/H$_3$PO$_4$ (serving as the separator) was saturated with BMIM-NTF$_2$ electrolyte. Two AGcf electrodes (X cm×X cm) and the PVA/H$_3$PO$_4$ membrane were sandwiched together to get the symmetric supercapacitor.

## Electrochemical characterization of Aqueous symmetric supercapacitor

The electrochemical performance of the symmetric supercapacitor was evaluated in a two-electrode system with a CHI 660E electrochemical workstation. CV tests were conducted at different scan rates from 10 - 100 $mV \cdot s^{-1}$. GCD tests were conducted at current densities from 5 - 20 $mA \cdot cm^{-2}$. EIS measurements were conducted in the frequency range of 100 $kHz$ - 0.05 $Hz$ at a 5 $mV$ amplitude referring to open circuit potential. The charge– discharge cyclic stability of the symmetric supercapacitor was carried by CV tests at 100 $mV \cdot s^{-1}$.

# Calculation

## Single electrode calculation

Absolute and areal capacitances of the single electrodes were calculated according to the following equation (by GCD data):

$$C = \frac{I \cdot \Delta t}{\Delta V} \quad (1)$$

$$C_a = \frac{C}{S} \quad (2)$$

Where $C$ (F) is absolute capacitance of one electrode, $C_a$ (F·cm$^{-2}$) is areal capacitance, $I$ (A) is the constant discharging current, $\Delta t$ (s) is the discharging time, $\Delta V$ (V) is the potential window (excluding the IR drop) and $S$ (cm$^2$) is the surface area of electrode.

Specific capacitance of the single electrodes were calculated from the following equations:

$$C_m = \frac{C}{m} \quad (3)$$

Where $C_m$ (F·g$^{-1}$) is specific capacitance, and $m$ (g) is the mass of electrode.

## Symmetric supercapacitor device calculation

The calculation of specific capacitance ($C_m$, F·g$^{-1}$), areal capacitance ($C_m$, F·g$^{-1}$), and cell (device) capacitance ($C_{cell}$, F) is similar to the calculation above.

Energy density (E, Wh·Kg$^{-1}$) and power density (P, KW·Kg$^{-1}$) of the device were obtained from the following equations:

$$E = \frac{1}{2 \times 3.6} \cdot C_m \cdot V_{max}^2 \quad (4)$$

$$P = \frac{(V_{max} - V_{drop})^2}{4 \cdot ESR \cdot m} \quad (5)$$

Where $V_{drop}$ is the voltage drop between first and second points from its cut-off of discharge curve, and $ESR\ (\Omega)$ is the internal resistance of the device obtained by:

$$ESR = \frac{V_{drop}}{2 \cdot I} \quad (6)$$

**Fig:**

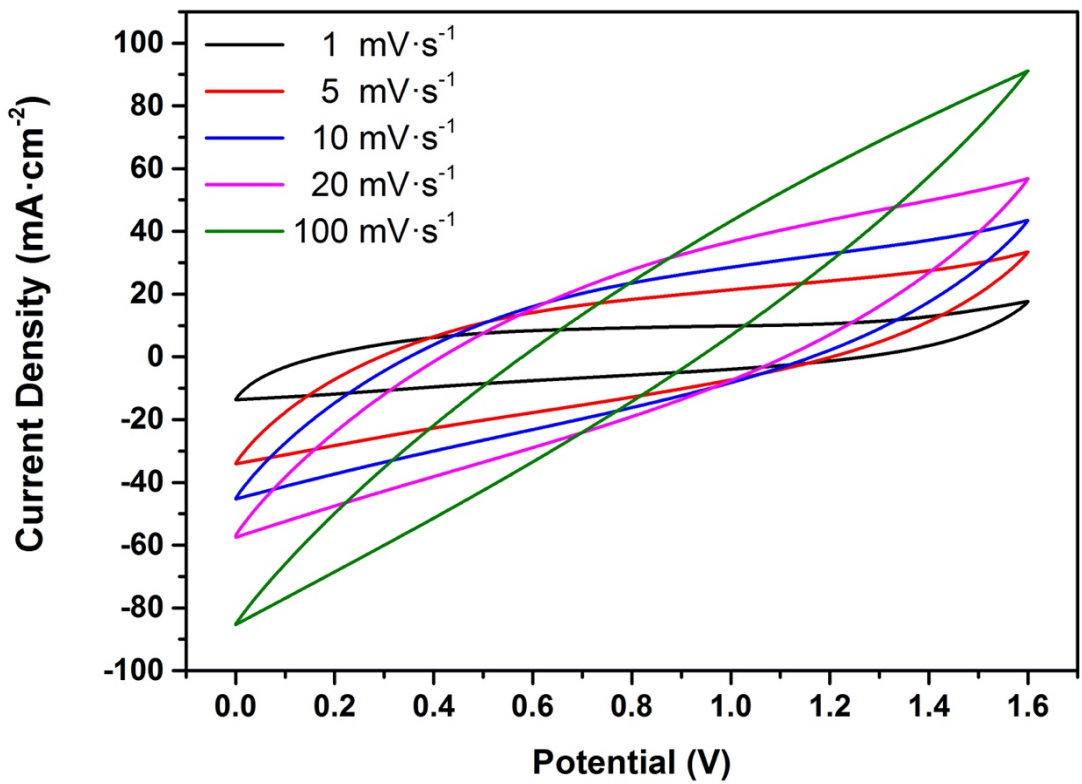

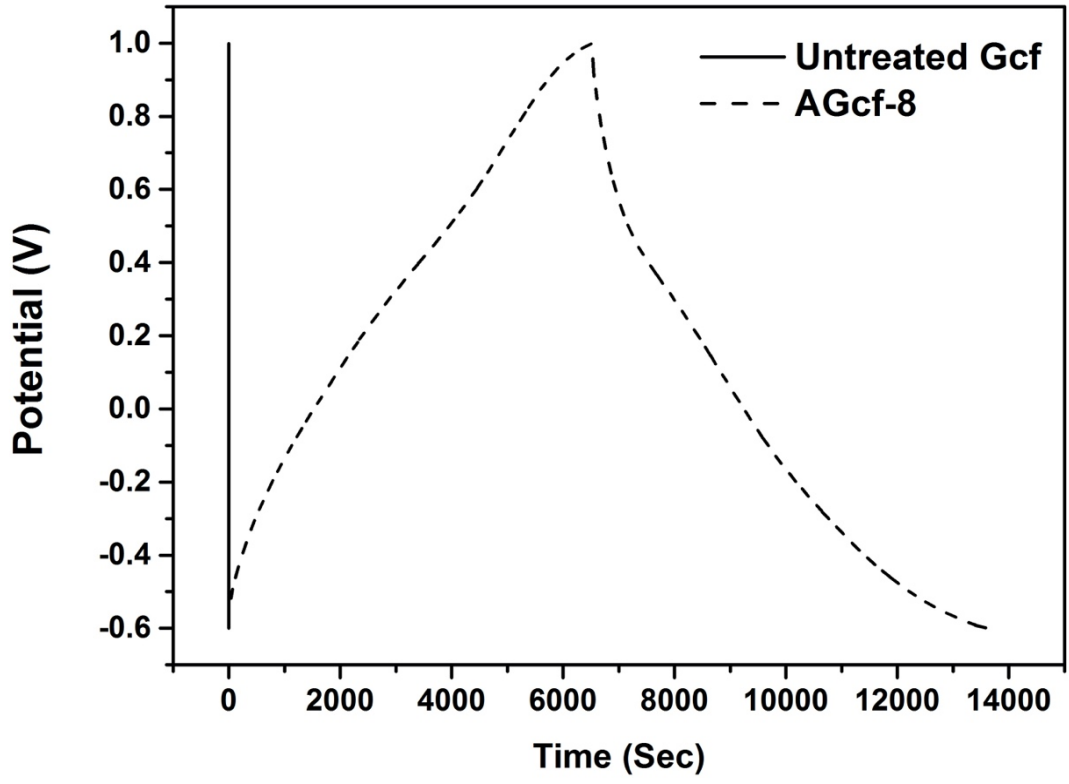

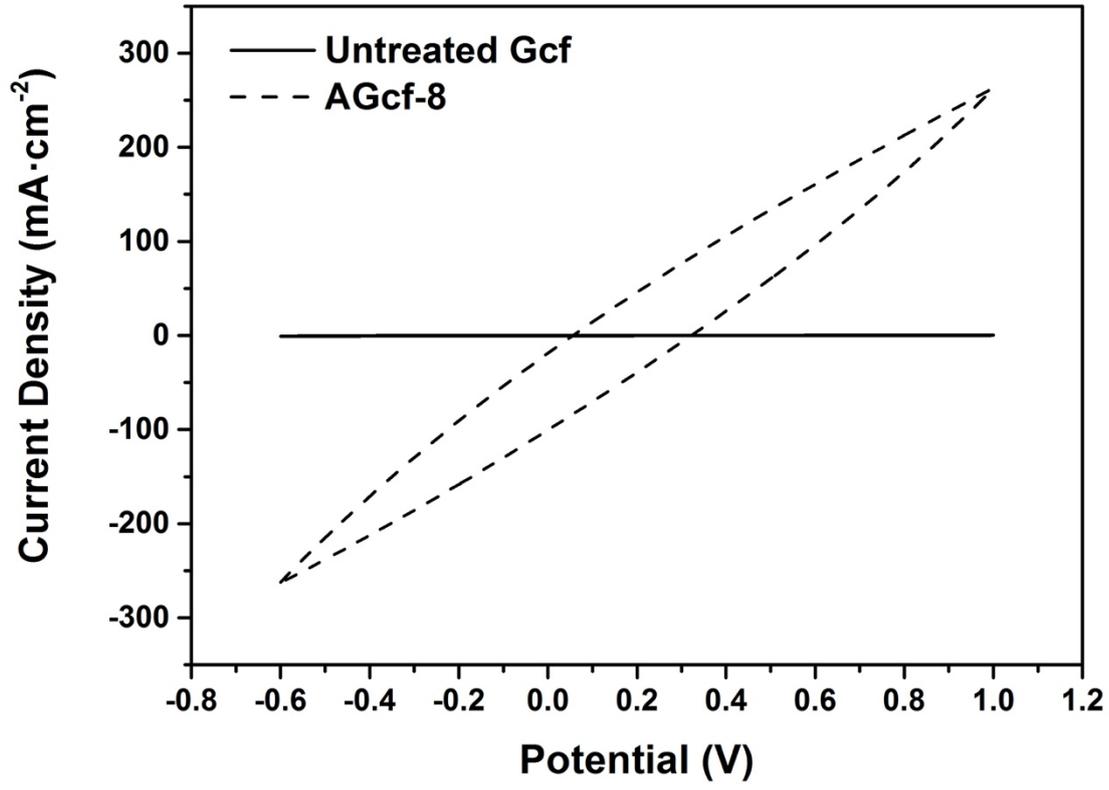

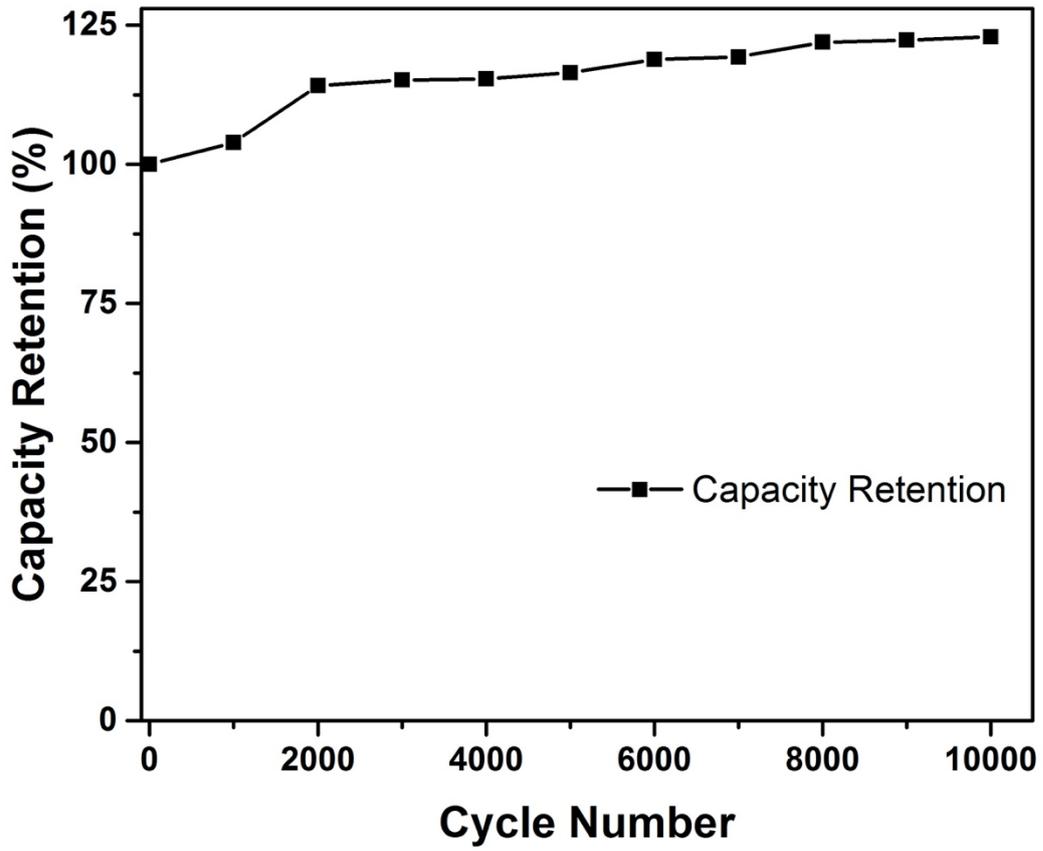